\documentclass[journal]{IEEEtran}

\usepackage{cite}      
\usepackage{graphicx}  
\usepackage[english]{babel}
\usepackage{subfigure} 
\usepackage{graphics}
\usepackage{float}
\usepackage{epsfig}

\begin{document}

\title{Laue Lens Development for Hard\\
X--rays ($>$60 keV)}

\author{D. Pellicciotta, F. Frontera, G. Loffredo, A. Pisa, K. Andersen, P. %
Courtois, B. Hamelin, V. Carassiti,%
\\ M. Melchiorri, S. Squerzanti%
\thanks{D. Pellicciotta is with Physics Dept, University of Ferrara, Ferrara, %
Italy.}%
\thanks{F. Frontera is with Physics Dept, University of Ferrara, Ferrara, %
Italy and with IASF, CNR, Bologna, Italy.}%
\thanks{G. Loffredo is with Physics Dept, University of Ferrara, Ferrara, Italy.}%
\thanks{A. Pisa is with Physics Dept, University of Ferrara, Ferrara, Italy.}%
\thanks{K. Andersen is with Institute Laue-Langevin, Grenoble, France.}%
\thanks{P. Courtois is with Institute Laue-Langevin, Grenoble, France.}%
\thanks{B. Hamelin is with Institute Laue-Langevin, Grenoble, France.}%
\thanks{V. Carassiti is with INFN, Ferrara, Italy.}%
\thanks{M. Melchiorri is with INFN , Ferrara, Italy.}%
\thanks{S. Squerzanti is with INFN , Ferrara, Italy.}%
}

\maketitle

\begin{abstract}
\noindent Results of reflectivity measurements  of mosaic crystal samples
of Cu (111) are reported. These tests were performed in the context of
a feasibility study of a hard X--ray focusing telescope for space astronomy 
with energy passband from 60 to 600 keV. The technique envisaged is that
of using mosaic crystals in transmission configuration that
diffract X-rays for Bragg diffraction (Laue lens). 
The Laue lens assumed has a spherical shape with focal length $f$. It is made of 
flat mosaic crystal tiles suitably positioned in the lens.  
The samples were grown and worked for 
this project at the Institute Laue-Langevin (ILL) in Grenoble (France), 
while the reflectivity
tests were performed at the X--ray facility of the Physics Department
of the University of Ferrara.
\end{abstract}

\begin{keywords}
X--ray, Astronomy, Telescope, Bragg, Laue, Crystals.
\end{keywords}

\section{Introduction}

\PARstart{T}{he} role of hard X--ray astronomy ($>$10 keV) is now widely recognized.
A breakthrough in the sensitivity of the hard X--ray telescopes, which today are based on
detectors that view the sky through (or not) coded masks (e.g.,\cite{Frontera97,Ubertini03}), 
is expected when focusing optics will be available also in this energy range. 
Focusing techniques are now in an advanced stage of development. The best technique
to focus hard X-rays with
energy less than about 70 keV, appears to be Bragg diffraction 
from multilayers. These are made of a set of bilayers, each consisting of
a low $Z$ together with a high $Z$ material, with graded thickness (see e.g. \cite{Joensen93}).
Above 70 keV, multilayer mirrors become inefficient (e.g. \cite{Fiore04})
and Bragg diffraction from mosaic crystals in transmission configuration (Laue geometry) 
appears to be the most efficient way to face the focusing problem at these energies. Mosaic
crystals and their diffraction properties have been known for many years; for a reference book
see \cite{Zac45}.
By mosaic crystal we mean a crystal made of microscopic perfect crystals (crystallites), 
with their lattice planes
slightly misaligned with each other around a mean direction according to a 
distribution function which can be approximated by a Gaussian:
%
%
\begin{equation}
	W(\Delta)=\frac{1}{\sqrt{2\pi}\eta}exp\left(-\frac{\Delta^2}{2\eta^2}\right)
\end{equation}
%
where $\Delta$ is the magnitude of the angular deviation of the crystallites from the mean 
direction, and $\eta$ is the standard deviation of the distribution.
The Full Width at Half Maximum (FWHM) of the Gaussian function defines the mosaic spread
$\beta \approx 2.35 \eta$ of the mosaic crystal. 
%
%
\begin{figure}[!b]
	\begin{center}
		\includegraphics[width=0.4\textwidth]{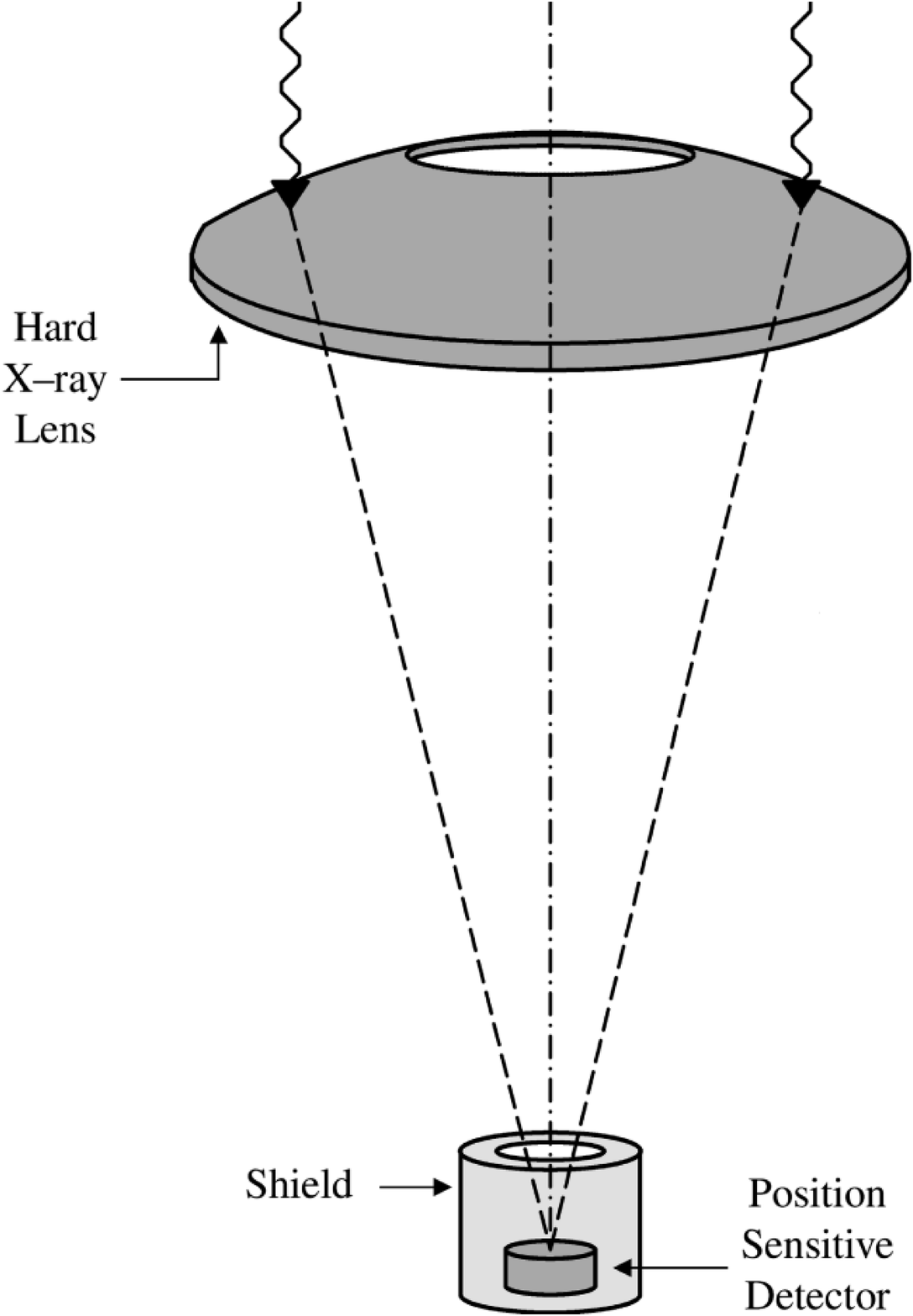}
		\caption{Pictorial view of a Laue lens (not to scale). The lens is 
			made of  mosaic crystal tiles in transmission configuration. 
		The impinging photons  are diffracted by the crystal tiles and focused
		in a small region centred in the focus of the lens where a 
		position-sensitive detector is positioned.} 
		\label{lens}
	\end{center}
\end{figure}
%

In general, the X--rays which impinge on a perfect crystal are diffracted  
according to the Bragg law:
%
%
\begin{equation}\label{BraggLaw}
	2d\sin\theta_B=n\frac{hc}{E}
\end{equation}
%
where $\theta_B$ (Bragg angle) is the angle between the lattice planes and the direction 
of both the incident and
diffracted photons, $2\theta_B$ is known as diffraction angle, 
$d$ (in \AA) is the distance between 
lattice planes, $n$ ($= 1,2, ..$) is the diffraction order, $E$ (in keV) is the 
photon energy and $hc=12.4$~keV$\cdot$\AA. 
In the case of a mosaic crystal, thanks to its mosaic $\beta$, when a polychromatic 
parallel beam of X--rays  impinges on it with mean Bragg angle $\theta_B$, photons in a bandwidth
%
%
\begin{equation}\label{DeltaE}
	\Delta E = E \, \beta / \tan \theta_B
\end{equation}
%
are diffracted by the crystal.

The goal of our project is to develop a broad band hard X--/gamma--ray 
($>$60~keV) focusing telescope devoted to the study of the continuum 
emission from celestial sources.  
The technique envisaged is  that of using mosaic crystals in transmission 
configuration and the telescope is designed to be made of mosaic crystal tiles
(Laue lens), where the reflected X--ray beam  emerges 
from the surface opposite to the crystal front surface. 
Laue lenses devoted to the study of the 
gamma--ray emission from nuclear lines, and thus with a relatively narrow 
energy passband,  have been already developed and tested 
(see, e.g., \cite{Vonballmoos04,Halloin04}).

\section{Summary of the lens main properties}

Results of the feasibility study of our lens have already been reported 
\cite{Pisa04}. Here we summarize the most relevant results.

\indent The lens of our project has a spherical shape (see Fig.~\ref{lens}) with radius 
$R$ and focal length $f = R/2$. In the lens focus a detector is positioned.
The detector is required to have a high detection efficiency in the lens passband 
(see below) and a position sensitivity proportioned to the imaging capabilities
of the lens. 

\begin{figure}[!t]
	\begin{center}
  	\includegraphics[width=0.5\textwidth]{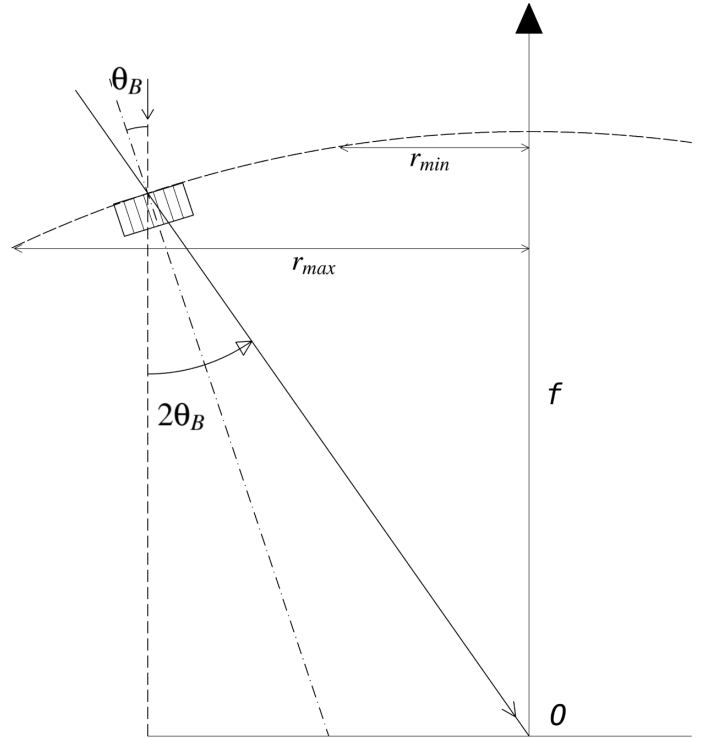}
		\caption{Scheme of the Bragg diffraction in a Laue lens.
		The X--ray photons in a given energy band $\Delta E$ around $E$, which impinge on the lens 
                 parallel to the $z$ axis (lens axis), are diffracted only by those mosaic
                 crystals oriented in such a way as to satisfy Eq.~\ref{DeltaE}. The photons 
                 with centroid energy $E$ which hit the center of these crystals 
			are focused in the lens focus (point $O$), $r_{max}$ and $r_{min}$
		define the innermost and outermost radius of the lens surface, respectively.}
		\label{fig:CrysPos}
	\end{center}
\end{figure}

The crystal tiles 
are assumed to have their lattice planes perpendicular to their
front surfaces (see Fig.~\ref{fig:CrysPos}), to be flat and
with small square cross section ($\sim 1\times1$~cm$^2$) 
in order to best approximate the spherical profile. 

Consistent with the Bragg diffraction principles above summarized, 
the photons in a band $\Delta E$ around a given  energy E which impinge on the lens 
parallel to the instrument axis ($z$ axis in Fig.~\ref{fig:CrysPos}) are diffracted 
only by those crystals which have their lattice planes oriented in such a way as
to satisfy  Eq.~\ref{DeltaE}, where $\theta_B$ is the average Bragg angle 
(see Eq.~\ref{BraggLaw}) of these crystals (see also Fig.~\ref{fig:CrysPos}).
Photons with the centroid energy $E$ are diffracted from their initial 
direction (see Fig.~\ref{fig:CrysPos}) by an angle $2\theta_B$, and are focused at
point $O$ of the focal plane if they impinge on the center of the crystal tile 
front surface, as in Fig.~\ref{fig:CrysPos}.

%
%
%
The lens external diameter increases with $f$, while its surface 
approximately increases with $f^2$.

\begin{figure}[!tb]
	\begin{center}
		\includegraphics[angle=-90,width= 0.5\textwidth]{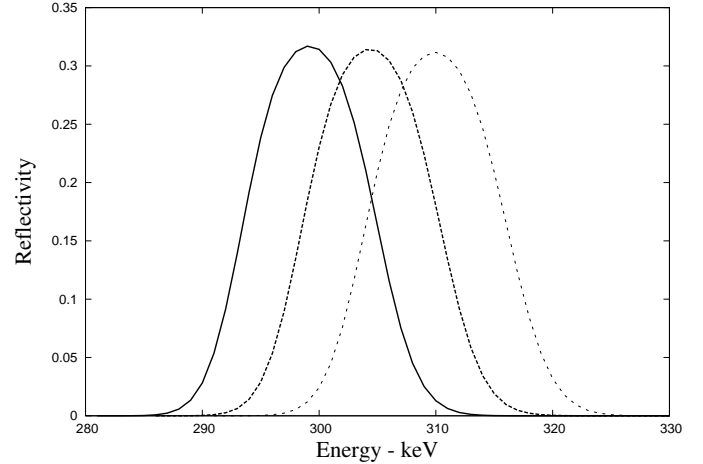}
		\caption{An example of the reflectivity profile of three contiguous 
		crystals along an Archimedes' spiral.} 
		\label{f:refl}
	\end{center}
\end{figure}

The mosaic crystal tiles are disposed in the lens according to an Archimedes' 
spiral (see Fig.~\ref{lens_tiles}). Thanks to this disposition 
the centroid energy of the bandwidth $\Delta E$ (Eq.~\ref{DeltaE}) changes in a very 
smooth manner along the spiral. Fig.~\ref{f:refl} shows the reflectivity 
profiles of 3 contiguous crystals along the spiral curve. As a consequence of
the Archimedes' spiral disposition, the effective area of 
the lens, defined as the lens geometric area projected in the focal plane times 
the mean reflection efficiency, smoothly changes with energy, apart from jumps due 
to the contribution from higher diffraction orders (see Fig.~\ref{A_eff}).

%
%

The nominal energy passband of the Laue lens ($E_{min},\,E_{max}$) is related to the 
range of diffraction angles ($\theta_{max}$, $\theta_{min}$) that are covered by 
the lens due to its spherical geometry, and thus (see Fig.~\ref{fig:CrysPos}) to 
the outer and inner radii $r_{max}$ and 
$r_{min}$ of the spherical segment. From the Bragg law, with simple calculations, 
it can be shown that
%
%
\begin{equation}
	E_{min}=\frac{hc}{2d \sin{\theta_{max}}} \approx \frac{hc f}{d~r_{max}}
\end{equation}
%
%
%
\begin{equation}
	E_{max}=\frac{hc}{2d \sin{\theta_{min}}} \approx \frac{hc f}{d~r_{min}}.
\end{equation}
%

\section{Mosaic Crystal Choice and Sample Test Results}

The reflection efficiency of the mosaic crystals, according to the theoretical 
model by \cite{Zac45}, is given by the following equation:
%
%
\begin{equation}
	R(\Delta,E)=\frac{I_d(\Delta,E)}{I_0}=\frac{1}{2}(1-e^{-2\sigma T})e^{-\mu \frac{T}{\gamma_0}}
	\label{e:refl}
\end{equation}
%
where $I_0$ is the intensity of the incident beam, $I_d(\Delta,E)$ is the intensity of 
the diffracted beam,
$\mu$ is the absorption coefficient per unit length at energy $E$, $\gamma_0$ 
is the cosine of the angle between the direction of the photons and the normal 
to the crystal surface, $T$ is the thickness of the mosaic crystal,
and $\sigma$ is given by:
%
%
\begin{equation}
	\sigma=\sigma(E,\Delta)=W(\Delta)Q(E)
\end{equation}
%
where
%
%
\begin{equation}
	Q(E)=\left|\frac{r_eF}{V}\right|^2\lambda^3\frac{1+\cos^2(2\theta_B)}{2\sin2\theta_B}
\end{equation}
%
in which $r_e$ (= 2.815$\times10^{-5}$\AA) is the classical electron radius, $F$ is 
the structure factor, $V$ is the volume of the crystal unit cell, $\lambda$ is the 
photon wavelength corresponding to the energy $E$, and $\theta_B$ is the
Bragg angle. 

As can be seen from Eq.~\ref{e:refl}, the mosaic reflectivity depends on various 
parameters, including the crystal material and the mosaic spread.  
As a result of our feasibility study \cite{Pisa04}, we have determined, 
in addition to the best lens shape and disposition of the crystals in the lens 
as discussed above, the best candidate materials and their required mosaic 
properties for photons in the energy range of interest. 

%
%
\begin{figure}[!tb]
	\begin{center}
		\includegraphics[width= 0.22\textwidth]{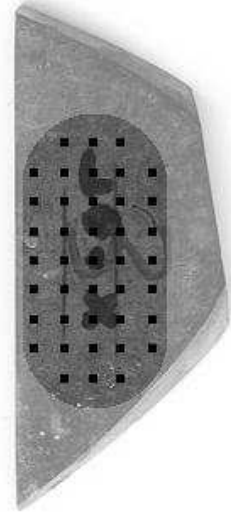}
		\caption{Picture of the front surface of a 2.5 mm thick crystal 
		sample of Cu (111), that was tested with a pencil beam 
            of polychromatic photons with $1.2 \times 1$~mm$^2$ 
		cross section. The  grid of the crystal surface positions hit by the beam is 
            shown. Lattice planes are perpendicular to the front surface and parallel to 
            the grid columns (vertical). The shady structures in the middle of the
		crystal are superficial inscriptions.} 
		\label{xtal}
	\end{center}
\end{figure}

Unfortunately the technological development of mosaic crystals is 
still in its infancy and only a few materials are available in a mosaic structure 
with the desired properties. Fortunately, for one of the best 
candidate materials (Copper), the technology has been recently developed 
at the Institute Laue-Langevin (ILL) in Grenoble (France).

%
%
\begin{figure}[!b]
	\begin{center}
		\includegraphics[width=0.5\textwidth]{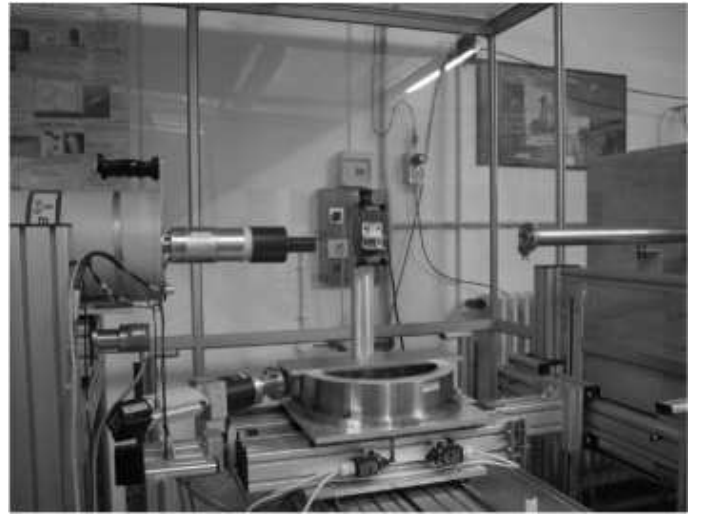}
		\caption{The X--ray facility of the Physics Department of the University of Ferrara
		and its main components. 
		On the left side: Ortec HPGe detector; in the center: sample holder;
		on the right: terminal collimator of the X--ray photon beam.} 
		\label{facility}
	\end{center}
\end{figure}

%
%
\begin{figure}[!t]
	\begin{center}
		\includegraphics[width=0.35\textwidth]{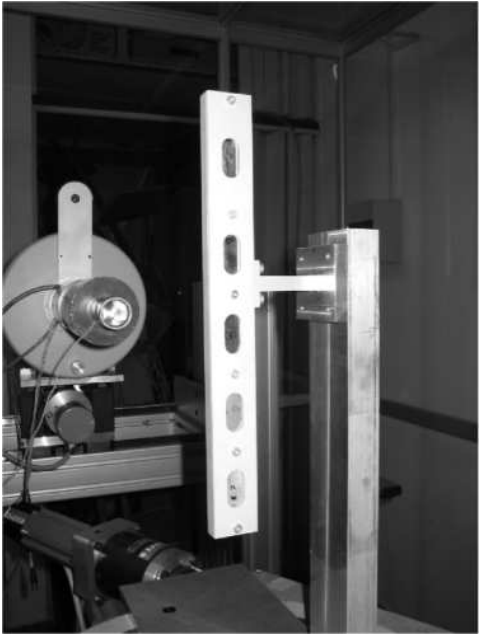}
		\caption{Holder and crystal samples used for the reflectivity tests.} 
		\label{holder}
	\end{center}
\end{figure}

In order to test the quality of the produced material and its
compatibility with the Laue lens constraints, some samples 
of Cu (111) of different thickness and mosaic spread, with their lattice
planes perpendicular to the crystal front surfaces,  have been
tested, and the reflectivity results compared with the expectations
of the mosaic crystal diffraction theory (Eq.~\ref{e:refl}). 

We concentrated our measurements on crystal samples with thickness 
in the 2-4 mm range, which is optimum for 
our lens (upper energy threshold of 600 keV), and with a mosaic spread below 6 arcmin. 
This upper limit is the maximum acceptable for low focal lengths (a few meters), 
given that the spread affects the focal spot size, which increases with
the focal length and affects the telescope sensitivity through the
focal plane detector background under the spot. 

The picture of the front surface of one of these samples is shown in Fig.~\ref{xtal}.  
The tests  were performed at the X--ray facility of the Physics Department of the 
University of Ferrara.
The Ferrara X--ray facility (see Fig.~\ref{facility}) allows the use of either 
polychromatic or monochromatic 
X--ray beams in the energy range from about 10 keV to 140 keV. This facility is
currently used for various types of measurements, from the calibration or test of hard
X--ray detectors to the test of hard X--ray mirrors. A description of the facility 
for the calibration of the JEM--X experiment ($\!\!\!$\cite{Lund03}) aboard the 
INTEGRAL gamma--ray astronomy satellite now in orbit, and the project for
the expansion of 
this facility now in progress, can be found elsewhere ($\!\!$\cite{Loffredo03,Loffredo05}).

Using a polychromatic pencil beam of hard X--ray photons with $1 \times 1.2\,$ mm$^2$ 
cross section, we measured the reflectivity of the samples 
at a grid of points over the crystal front surface (see Fig. \ref{xtal}). 
The tested crystals (see Fig.~\ref{holder}) were oriented
in order to get a mean Bragg angle $\theta_B$ with respect to the
polychromatic pencil beam corresponding to an energy of 90-100 keV. 
The crystal samples could be rotated in order to change, if desired, the 
energy centroid of diffracted photons (see Eq.~\ref{BraggLaw}).
The intensity of the direct and diffracted beams were measured with 
a portable cooled HPGe detector by Ortec (energy resolution better than 1\% at 122 keV).

Given the polychromatic nature of the beam, after subtraction of the 
detector background, the ratio between the diffracted and
the direct beam gives the X--ray reflectivity profile of the crystal cross section
hit by the photons. The X--ray reflectivity curve at different points of 
the crystal surface was measured by moving the crystal holder in a plane 
perpendicular to the pencil beam.


%
%
\begin{figure}[!t]
\begin{center}
	\includegraphics[width=0.5\textwidth]{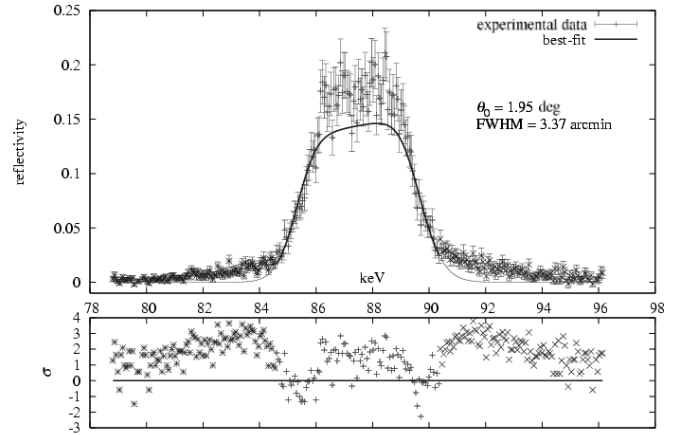}
	\caption{{\it Top panel}: Measured reflectivity curve of one of the crystal
      samples when the pencil beam hits the center of its front surface 
      (see position grid in Fig.~\ref{xtal}), before the chemical 
	removal of an external layer.
	Also shown is the best fit curve obtained by fitting  the 
	model function (Eq.~\ref{e:refl}) to the data. The best fit 
	parameters of the model (Bragg angle of diffraction and FWHM of the mosaic spread)
	are shown. In the fit, the thickness of the crystallites was frozen (0.02 $\mu$m). 
	{\it Bottom panel}: Residuals of the data to the model in units of standard
	deviations.} 
	\label{fit_1}
\end{center}
\end{figure}

%
%
\begin{figure}[!b]
	\begin{center}
		\includegraphics[width=0.5\textwidth]{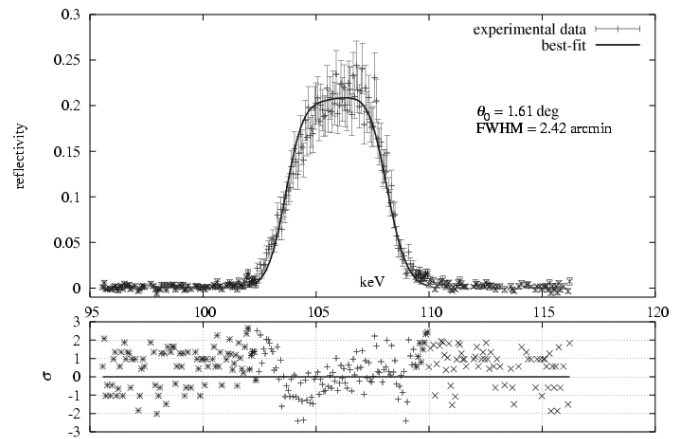}
		\caption{
		{\it Top panel}: Measured reflectivity curve of the same
		crystal sample of Fig.~\ref{fit_1} 
		after the chemical removal of an external layer of 0.1 mm thickness. 
		Also shown is the best fit curve obtained by fitting  the 
		reflectivity model function (Eq.~\ref{e:refl}) to the data. The best fit 
		parameters of the model (Bragg angle of diffraction and FWHM of the 
		mosaic spread) are shown. In the
		fit, the thickness of the crystallites was frozen (0.02~$\mu$m).
	  {\it Bottom panel}: Residuals of the data to the model in units of standard
		deviations.} 
		\label{fit_2}
	\end{center}
\end{figure}
For each sample, two sets of measurements were performed: one set 
before the removal of a thin layer of material from the crystal surfaces, and the
other set after this removal.

All the reflectivity curves were fit with the reflectivity 
model function of mosaic crystals in a Laue configuration (Eq.~\ref{e:refl}).
We used the CERN code MINUIT to perform the fit to the data. 
The free parameters of 
the fit were the Bragg angle, the mosaic spread $\beta$ 
and the thickness $t_0$ of the crystallites. Given that the last parameter
was not well constrained, it was fixed at 0.02 $\mu$m.

\begin{figure}[!t]
	\subfigure[]{\includegraphics[height=0.5\textwidth, angle=-90]{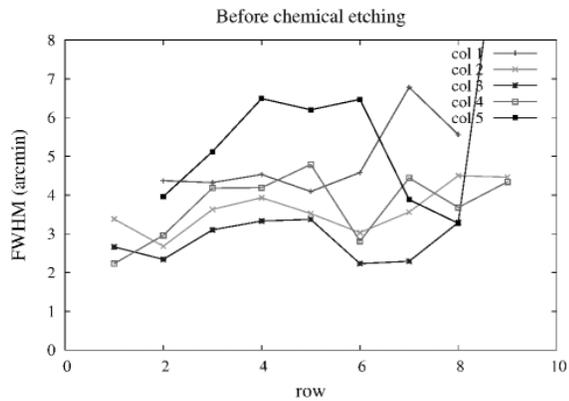}
	\label{fwhmb}}
	\vfil
	\subfigure[]{\includegraphics[height=0.5\textwidth, angle=-90]{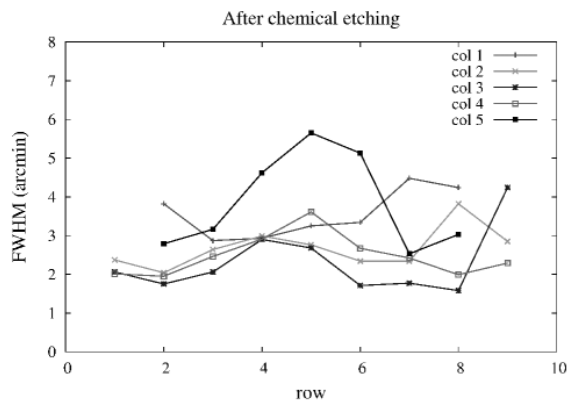}
	\label{fwhma}}
	\caption{Distribution of the measured mosaic spread of the crystal of 
	Fig.~\ref{fit_1} as a function of the grid position hit by the pencil beam.
	{\it Panel (a)}: before the removal of the external layer of 0.1 mm thickness; 
	{\it panel (b)}: after the removal of  the 0.1 mm thickness.}
	\label{fwhm}
\end{figure}

Fig.~\ref{fit_1} is an example of the measured reflectivity curve of a crystal sample 
of Cu (111) before the removal of the external layer, when the X--ray beam hits the
crystal  at center of the front surface (see Fig. \ref{xtal}).
The best fit model, superposed to the data, is also shown in Fig.~\ref{fit_1} along with  
its best fit parameters.
As can be seen, the measured reflectivity profile shows two 
side wings, which the best fit reflectivity model is unable to describe. 
Due to these wings, the reflectivity model (Eq.~\ref{e:refl}) is unable to
achieve the flat top of the measured profile.
This feature is found  at all tested points of the crystal surface.
For the same crystal sample, Fig.~\ref{fwhmb} shows the distribution 
of the mosaic spread
as a function of the row number of the grid of points tested, for different
columns of the grid (see map in Fig.~\ref{xtal}).
%
%

After the removal of a thin layer of material (0.1 mm) from the crystal, 
the reflectivity model fits  the data significantly better. As an example,
in Fig.~\ref{fit_2} we shows the corresponding reflectivity curve of the same
sample of Fig.~\ref{fit_1} after the removal of the external layer. 
As can be seen, the side wings are weakened and the reflectivity model
well describes the flat top data. These features are repeated at almost any point 
of the crystal surfaces tested. Fig.~\ref{fwhma} shows the corrisponding 
distribution of the mosaic spread for the same crystal sample.

A likely explanation for the behaviour of the crystal reflectivity 
before the removal of the thin layer is that the slicing of the crystal ingot (by
means of electro-erosion) to get the samples, perturbs the superficial properties 
of the crystal tile, 
giving rise to a reflectivity component with a larger mosaic spread which superposes 
on that of the crystal
bulk. The fact that the fit is still not so good even after the removal of the 
external layer could also mean that the theory of mosaic crystals is an 
approximate description of the real crystals.
In spite of this, the above results give strong support to our project and open new 
possibilities for focusing high energy X--ray photons.

\section{The Laue Lens Prototype}

\begin{figure}[!t]
	\begin{center}
		\includegraphics[width=0.4\textwidth]{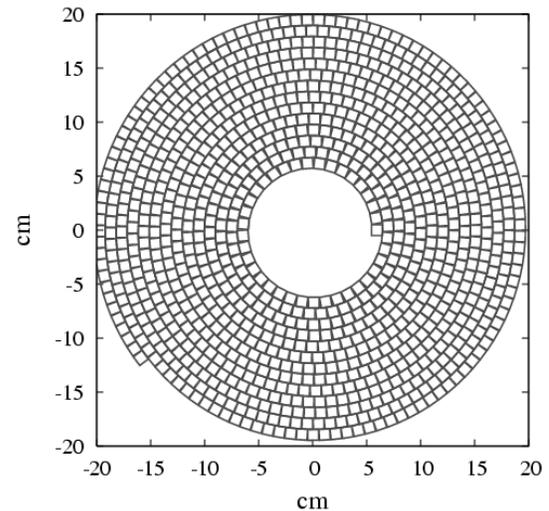}
		\caption{Configuration of the Laue lens prototype.  
		The crystal tiles are disposed
		along an Archimedes' spiral (see text).} 
		\label{lens_tiles}
	\end{center}
\end{figure}

On the basis of the test results obtained, we are developing a Laue lens prototype 
model (PM) made of 500 Cu (111) crystal tiles (see Fig.~\ref{lens_tiles}).  
The area of the front surface of all crystals is 15$\times$15 mm$^2$, 
while the crystal thickness is 2 mm for 415 tiles, and 4 mm for the others.  
The PM size is shown in Fig.~\ref{lens_tiles}, its
focal length is 210 cm, and the nominal energy passband is from 60 to 200 keV.

%
%
%
%
\begin{figure}[!ht]
	\begin{center}
		\includegraphics[width=0.5\textwidth]{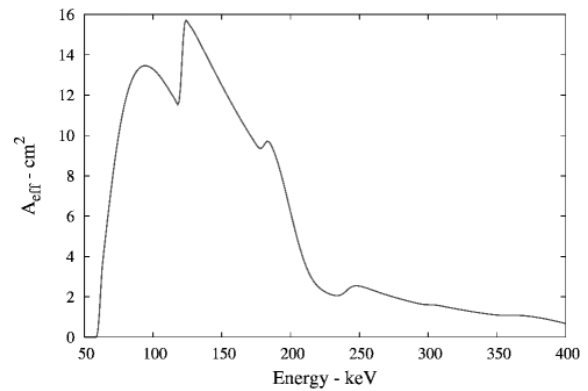}
		\caption{Expected effective area of the PM as a function of
		the photon energy, assuming a mosaic spread of 5 arcmin.
		The nominal energy passband of the PM is from 60 to 200 keV. The jumps
		in the effective area are due to the higher orders of diffraction
		from the outermost crystals.} 
		\label{A_eff}
	\end{center}
\end{figure}

The expected effective area of the PM as a function of the photon energy is shown in
Fig.~\ref{A_eff}, for an assumed mosaic spread of 5 arcmin.
As discussed above, the spiral configuration allows a smooth behaviour of
the lens effective area with energy, if we exclude the jumps at 120, 180 and 240 keV
which are due to the higher orders of diffraction from the outermost crystals.

If  the laboratory tests of the prototype are satisfactory, the prototype will 
also be tested aboard a balloon flight experiment.

\section{Conclusions}

We have reported on test results obtained from mosaic crystal samples of
Cu (111) produced at the Institute Laue-Langevin of Grenoble (France). 
According to a theoretical feasibility study of a Laue lens performed by
us ($\!\!$\cite{Pisa04}), these crystals are among the best materials to be used
for maximizing the capabilities of a Laue lens for hard X--rays ($>$ 60 keV)
for space astronomy. The sample test results obtained are consistent with the
theoretical expectations if a thin external layer is removed from the crystal
samples cut from a grown ingot of mosaic crystal of Copper. The mosaic spread
of the crystal samples shows a dependence on the crystal 
surface point hit by the X--ray photons. A better quality control of the
crystal production is desirable.
These results encouraged us to develop a lens prototype now in progress. 

\section*{Acknowledgements}
This research was supported by the Italian Space Agency ASI. We wish to thank
John B. Stephen for useful comments and a careful reading of the manuscript.

\end{document}